\documentclass[aps,prb,showpacs]{revtex4}
\usepackage{amsmath}
\usepackage{amssymb}
\usepackage{epsfig}

\begin{document}
\title{Pumping and electron-electron interactions}
\author{P. Devillard$^{1,2}$, V. Gasparian$^3$, and T. Martin$^{1,4}$}
\affiliation{$^1$ Centre de Physique Th\'eorique de Marseille
        CPT, case 907, 13288 Marseille Cedex 9 France}
\affiliation{$^2$ Universit\'e de Provence, 3, Place Victor Hugo, 
13331 Marseille cedex 03, France}
\affiliation{$^3$ Department of Physics, California State University,
Bakersfield, CA, USA}
\affiliation{$^4$ Universit\'e de la M\'editerran\'ee, 13288 Marseille Cedex 9, France}

\begin{abstract}
We consider the adiabatic pumping of charge through a mesoscopic 
one dimensional wire in the presence of electron-electron interactions.
 A model of static potential in-between two-delta potentials is used to obtain exactly 
the scaterring matrix elements, which are renormalized by the interactions.
 Two periodic drives, shifted 
one from another, are applied at two locations of the wire in order to 
drive a current through it in the absence of bias. Analytical expressions 
are obtained for the pumped charge, current noise, and Fano factor in different regimes. 
This allows to explore 
pumping for the whole parameter range of pumping strengths.
 We show that, working close to a resonance is necessary to have a comfortable window
of pumping amplitudes where charge quantization is close to the optimum value: a single
 electron charge is transferred in one cycle. Interactions can improve the situation, 
the charge $Q$ is closer to one electron charge and noise is reduced, following a
 $Q \,(e-Q)$ behavior, reminiscent of the reduction 
of noise in quantum wires by $T \, (1-T)$, where $T$ is the electron transmission 
coefficient.
 For large pumping amplitudes, this charge vanishes, noise also decreases but slower than the charge.
\end{abstract}
\maketitle

\section{Introduction}
\label{sec:introduction}

The suggestion that electrons can be supplied one by one by a mesoscopic 
circuit has been proposed over two decades ago \cite{thouless}.
Instead of applying a constant bias voltage to the system, it is possible
to supply a.c. gate voltages which perturb the system periodically.
Under certain conditions, the charge transferred from one lead to the other,
during one period, can be almost quantized.
Adiabatic pumping of electrons could in principle be used in 
future nanoelectronics schemes based on single electron transfer, 
and it also has applications to 
quantum information physics. Over the years, 
theoretical approaches to this adiabatic pumping based 
on scattering theory have become available 
\cite{buttiker_thomas_pretre,brouwer,MB02}. 
These situations typically describe mesoscopic systems which are large enough, 
or sufficiently well connected to leads that electronic interactions (charging effects
for instance) can be discarded. 
Scattering theory has been
applied\cite{MB02} to calculate both the charge and noise in systems
in the absence of electron-electron interactions.

On the experimental scene, Coulomb blockade effects have 
been successfully exploited to achieve pumping with isolated quantum dots \cite{kouvenhoven}. 
To our knowledge, pumping experiments which are not entirely based on Coulomb blockade, 
where the shape of the electron wave functions is modified in an adiabatic
drive are rather scarce.  A recent study \cite{marcus} has dealt with 
the transport through an open quantum dot where such interactions are minimized. 

Besides Coulomb blockade physics, the effect of electron-electron 
interactions in conductors with reduced dimensionality have been 
discussed by several authors. The case of strong interactions
 in a one dimensional
quantum wire was presented in Ref. \onlinecite{sharma_chamon}, 
using Luttinger liquid theory.  Alternatively, Ref. \onlinecite{rao}
discussed the opposite limit, where the effect of weak interactions
can be included in a scattering formulation of pumping using renormalized
transmission/reflection amplitudes\cite{DGM94}. 
However, the results for the pumped charge remain mostly numerical
in this work.

The conditions under which pumping amplitude and interactions
must be tuned to achieve quantized pumping are not obvious. Many
physical parameters enter this problem, such as 
the amplitude of the pumping potentials, the phase difference between these, 
the possibility of a constant offset on these potentials, the overall
conductance of the unperturbed structure, and to what extent
the strength of electron-electron interactions play a role. 
Analytical results on this issues are highly desirable, as well as
information about the noise.

With regard to the experiment of Ref. \onlinecite{marcus}, there is clearly a need for
further understanding the role of weak interactions in such mesoscopic 
systems in the presence of pumping. 
The purpose of the present work goes in this direction, in the sense that 
we provide analytical expressions for the pumped charge and the noise for 
a one dimensional wire in the presence of interactions. 
This allows us to explore all pumping regimes\cite{GAO05} (weak to strong pumping)
and to determine in which manner and to what extent the pumped charge can help 
to achieve single electron transfer.        
Besides addressing the question of the ideal 
conditions for good charge quantization, we shall establish relationships
 between charge and noise in different regimes. 
For concreteness, a two-delta potentials model will be used 
and interactions will be added on top of it.    

\section{Pumped charge and noise}

\subsection{Adiabatic pumping in non interacting systems}

Here, we recall the formula which was established for the  
charge transferred during the single period of an adiabatic pumping cycle
through a quasi one-dimensional system. The system is in general described by
a potential $V(x)$ containing two internal parameters which 
are modulated periodically. The time dependence is assumed to be sufficiently 
slow so that, although the scattering matrix depends on time, its variations
are minute when an electron is scattered in the mesoscopic wire.

At finite temperature, the pumped current reads\cite{Polianski02}
\begin{eqnarray}
Q \, =\, e \int_0^{{2 \pi \over \omega}} 
dt \int f(E) \, Tr \bigl\lbrace S^{\dagger}(E,t) \sigma_z S(E,t) - I \bigr\rbrace
 {dE \over 2 \pi}.
\end{eqnarray}
where $S(E,t)$ is the Wigner transform of the scattering matrix $S(t,t^{\prime})$,
 and to, a good approximation  the scattering matrix for the problem with 
the potential frozen. $\sigma_z$ is the ususal Pauli matrix, $I$ the identity matrix and
 $f$ the Fermi-Dirac function.
\begin{eqnarray}
S(E,t) \, =\, \int_{- \infty}^{\infty} e^{-i E (t-t^{\prime})} S(t,t^{\prime}) dt^{\prime}.
\end{eqnarray}

The pumping potential will generate sidebands at $E \pm \hbar \omega$ 
and we assume that temperature is much smaller than  $\omega$,
 i.e. $k_B T \ll \hbar \omega$, so that we can approximate 
the Fermi function by a step function. In fact temperature dependence
 occurs in two places in this problem. First in the Fermi function
 and second, the scattering matrices elements depend on the temperature because
 of the renormalization due to the interactions (see next chapter). Formulas
 for averaged current and zero frequency noise can be carried out using results
 of the literature in the ``zero temperature'' formalism, except that the
 scattering matrix elements are in fact temperature dependent.

The pumped charge reduces to a time integral over a pumping cycle\cite{
buttiker_thomas_pretre,rao}
\begin{eqnarray}
Q \, =\, {e \over 2 \pi} \int_0^{2 \pi} 
Im \Biggl\lbrack \Biggl( {\partial s_{11}\over \partial X} s_{11}^*
 + {\partial s_{12} \over \partial X} s_{12}^* \Biggr) {dX \over dt} + 
(X \leftrightarrow Y) \Biggr\rbrack \, dt,
\label{Qequation}
\end{eqnarray}
where $s_{1i}$ ($i=1,2$) are the elements of the scattering matrix $s(E)$:

\begin{equation}
{s(E)}=e^{i\phi}\left(\begin{array}{cc}
-i \sqrt R e ^{i\theta} &\sqrt T \\
\sqrt T& -i \sqrt R e^{-i\theta}
\end{array} \label{sm}\right),
\end{equation}
where $\phi$ is the phase accumulated in a transmission event and $\theta$ is the phase
 characterizing the asymetry between the reflection from the left-hand-side and from 
 the right-hand-side of the potential. Conservation of probabilities
 imposes $R+T=1$. We assume the quantities $\sqrt{R}$, $\sqrt{T}$,
 $\theta$ and $\phi$ to be functions
 of the Fermi energy $E_F$ and of
 the external time-varying parameters $X(t)$ and $Y(t)$.
 
\subsection{Inclusion of weak interactions}

In the case of weak interactions, the transmission and
the reflection amplitudes $s_{12}$ and $s_{11}$ can be calculated 
in the presence of Coulomb interaction via
 a renormalization procedure\cite{DGM94}. 
High energy scales above a given cutoff
are eliminated. The high energy cutoff is lowered progressively. 
The renormalization 
has to be stopped when the temperature becomes comparable to this cutoff. 
Finally, 
if $s_{12}^{(0)}$ and
 $s_{11}^{(0)}$ denote respectively, 
the transmission and reflexion  coefficient without interactions, $s_{12}$ and $s_{11}$
 can be expressed in the form
\cite{DGM94}:
\begin{eqnarray}
s_{12}&=&{s_{12}^{(0)} l^{\alpha}\over \sqrt{1+T_0(l^{2\alpha}-1)}}, \label{t1}\\
s_{11}&=&{s_{11}^{(0)}\over \sqrt{1+T_0(l^{2\alpha}-1)}}\label{r1},
\end{eqnarray}
where $l= k_B \Theta/W$; $\Theta$ is the temperature, $k_B$ the Boltzmann
 constant and $W$ the original bandwidth. 
$\alpha$ is a negative exponent related to
 the strength of the screened Coulomb interaction potential\cite{DGM94}. Specifically:
\begin{equation}
\alpha={V_c(2k_F)-V_c(0)\over 2\pi v_F},
\end{equation}  
with $V_c(q)$ the Fourier transform of the screened Coulomb potential at $q=2k_F$ and $q=0$, 
respectively, $V_c(0)$ is finite due to screening.
 $\alpha=0$ corresponds to the absence of electron-electron interactions. 
$T_0=|s_{12}^{(0)}|^2$
 represents the conductance of the wire in  units of $e^2/h$. 
>From now on, $Q$ will denote the pumped charge with
interactions and $Q_0$ without interactions. 
The integrand of Eq. (\ref{Qequation}) is therefore 
modified by the presence of the interactions.
 Thus, temperature dependence occurs through the renormalization of the $S$-matrix. 
We shall be interested in the regime where temperature is much lower than pumping frequency,
 $k_B \Theta \ll \omega$, as said before, 
but renormalization of $S$-matrix should not be too 
severe so that the renormalization of $S$-matrix
 still makes sense. At very low temperature, all barriers become almost opaque and
 bosonization is required \cite{sharma_chamon}, so typically $l^{\alpha} > 10^{-1}$,
 that is $k_B \Theta \gg  W \, 10^{1 / \alpha}$. 
For example, for nanotubes having $\alpha$ around $-0.3$,
 this gives $\hbar \omega \gg\ k_B \Theta \gg \, 10^{-3} W$.

\subsection{Two-delta potentials model and pumped charge}

Consider the pumping charge $Q$ transferred during a single period through
 a 1D chain of an arbitrary potential shape. Let two parameters of the system be modulated
 periodically. The single particle Hamiltonian reads
\begin{eqnarray}
H \, =\, {\hbar^2 k^2 \over 2 m} + V(x) + V_p(x,t),
\label{Hsingle}
\end{eqnarray}
where $V_p(x,t)$ is the time-dependent perturbation part of the arbitrary potential and
 has $\delta-$like potentials form
\begin{eqnarray}
V_p(x,t) \,= \, 2k_F X(t) \delta(x-x_i) + 2 k_F Y(t) \delta(x-x_f),
\label{definitionofVp}
\end{eqnarray}
with the amplitudes $X(t) = V_i(t)/2k_F$ , $Y(t) = V_i(t)/2k_F$,
 where $V_i(t)$ and $V_f(t)$ have periodic time evolution with
 the same period $t_0 = 2 \pi/\omega$ and $k_F$ denotes the Fermi wave-vector,. $V(x)$ 
is the static potential
 in between two $\delta$ potentials. Below, superscript indexes
 $(0)$ indicate non-interacting systems.

Using the known relation \cite{Fisherlee81}, 
\begin{eqnarray}
s_{\alpha \beta}^{(0)} \, =\, - \delta_{\alpha \beta} + 2 i k_F G^0(x_{\alpha},x_{\beta}),
\end{eqnarray}
where $G^0(x_{\alpha},x_{\beta})$ are the usual real space retarded Green's functions 
and the fact that the functional derivative of the Green's function 
$\delta G^0/\delta V(x)$ can be written as the product of 
two Green's functions\cite{GCB96}
\begin{eqnarray}
{\delta G^0 (x_{\alpha},x_{\beta})\over \delta V(x_j)} \, =\, G^0(x_{\alpha},x_j)
 G^0(x_j,x_{\beta}).
\end{eqnarray}
For the first bracket of Eq. (\ref{Qequation}), we get
\begin{eqnarray}
{\partial s_{11}^{(0)} \over \partial X} s^{(0) \, *}_{11} + 
{\partial s_{12}^{(0)} \over \partial X} s^{(0) \, *}_{12}
 \, =\, {\bigl(1+s_{11}^{(0)}\bigr) \bigl(1+s_{11}^{(0) \,*}\bigr) \over 2 i k_F},
\label{firstbracket}
\end{eqnarray}
where we used the condition $s^{(0)}_{11} s^{(0) \, *}_{11} +
 s_{12}^{(0)} s_{12}^{(0) \,*} = 1$, $s_{12}^{(0)}$ and $s_{11}^{(0)}$ are
 the bare transmission and reflection amplitudes from
 the disordered system, without taking account the electron-electron interactions.
For the second bracket of Eq. (\ref{Qequation}), we get
\begin{eqnarray}
{\partial s_{11}^{(0)} \over \partial Y} s_{11}^{(0) \, *} + 
{\partial s_{12}^{(0)} \over \partial Y} s_{12}^{(0) \, *} \, =\,
{s_{12}^{(0)} \over 2 i k_F} \, \Bigl(s_{12}^{(0)} s_{11}^{(0) \,*} +
 s_{12}^{(0) \,*} (1+s_{22}^{(0)})\Bigr) \, =\,
{\vert T_0\vert^2 \over 2 i k_F}.
\label{secondbracket}
\end{eqnarray}
Here, we used the
 current conservation requirement $s_{21}^{(0)} s_{11}^{(0) \,*}
 + s_{12}^{(0) \,*} s_{22}^{(0)} =0$.
$s_{22}^{(0)}$ is the reflection
 amplitude from the right of the scaterrer and can be presented as
\begin{eqnarray}
s_{22}^{(0)} \, =\, 2 i k_F {\partial ln s_{12}^{(0)} \over \partial Y} -1.
\label{rprimeidentity}
\end{eqnarray}
Substituting Eq. (\ref{firstbracket}) and (\ref{secondbracket}) into Eq. (\ref{Qequation}),
 using the identity (\ref{rprimeidentity}), we finally arrive at
\begin{eqnarray}
Q_0 \, =\, -{e \over 2 \pi} \int_0^{{2 \pi \over \omega}} \Biggl(
{\partial ln s_{12}^{(0)} \over \partial X} {\partial ln s_{12}^{(0) \,*} \over \partial X} 
{dX \over dt} + T_0 {dY \over dt} \Biggr) \, dt.
\label{Q0sint}
\end{eqnarray}
Similar expression for $Q_0$, can be found in Ref. \onlinecite{avron}.

Using the same method for $Q$, we obtain, in the presence of interactions
\begin{eqnarray}
&Q& \, = \, Q_0 - {e (l^{2 \alpha} -1) \over 2 \pi}  \nonumber \\
&\times& \! \int_0^{{2 \pi \over \omega}} 
\Biggl\lbrack
 \biggl(
 Im \Bigl\lbrace {\partial \ln \, s_{12}^{(0)} \over \partial X} \Bigr\rbrace
 \, - 
\Bigl\vert{\partial \ln \,s_{12}^{(0)} \over \partial {X}}\Bigr\vert^2
\biggr) \, {d {X} \over dt}
\, + \,
\biggl( Im \Bigl\lbrace{\partial \ln \, s_{12}^{(0)} 
\over \partial {Y}}\Bigr\rbrace
 - T_0\biggr) \, {dY  \over dt}
\Biggr\rbrack
\, {T_0 \over 1 + T_0(l^{2 \alpha}-1)}\, dt. 
\label{Qsint}
\end{eqnarray}

For numerical simulations, we specialize to the case where the static potential 
$V(x)=0$ in Eq. (\ref{Hsingle}) and the time-dependent perturbations are two
 $\delta$-like potentials separated by a distance $2a$.
The expressions of $s_{12}^{(0)}$ and $s_{11}^{(0)}$ 
are needed, when the single particle Hamiltonian reads
\begin{eqnarray}
H \, = \, -\hbar^2 k^2/2m  + V_p(x,t).
\end{eqnarray}
The elements of the $S$-matrix in the absence of electron-electron interactions, are given by
\begin{eqnarray}
s^{(0)}_{11} &=& {\Bigl\lbrack ({\overline Y} - {\overline X}) sin(2 k_F a) 
- i 
 \bigl\lbrace 2 {\overline X} \, {\overline Y} sin(2 k_F a) + ({\overline X} 
+ {\overline Y}) cos(2k_F a) \bigl\rbrace \Bigr\rbrack \over D}, \label{equationsforsmatrix1} \\
s^{(0)}_{12} \, &=& \, {1 \over D},
\label{s12}
\end{eqnarray}

with
\begin{eqnarray}
D \, &=&\, \biggl(1 - 2 {\overline X} \, {\overline Y} sin^2(2 k_F a) + i \bigl\lbrack 
{\overline X} + {\overline Y} + {\overline X} \,{\overline Y} sin(4 k_F a) \bigr\rbrack
 \biggr),
\label{equationsforsmatrixlast}
\end{eqnarray}
and ${\overline X} = X/2$, same for ${\overline Y}$.
$s^{(0)}_{21} = s^{(0)}_{12}$ and $s^{(0)}_{22}$ is obtained by replacing ${\overline Y}$
 by ${\overline X}$ in $s^{(0)}_{11}$.

\subsection{Noise}

Ref. \onlinecite{avron} studied adiabatic quantum pumping in the context
 of scattering theory.
 Their goal was to derive under what conditions pumping could be achieved optimally,
 in a noiseless manner, with
 the assumption that the pumping frequency is small compared to the temperature.
 This enabled the authors to derive
 expressions not only for the pumped charge per cycle, but
 also for the pumped noise, the current-current correlation function, averaged
 over a time which is long compared to the period of the adiabatic drive, at zero
 frequency. Specifically, the noise is defined 
from the current-current time correlator:
\begin{eqnarray}
S(t,t^{\prime}) \, = \, {1 \over 2} \langle \delta I(t) \delta I(t^{\prime})  
+ \delta I(t^{\prime}) \delta I(t) \rangle,
\end{eqnarray}
with $\delta I \, = \, I - \langle I \rangle$. This correlator is then averaged over 
 $n_0$ periods of the pumping drive with $n_0$ large, and it is taken at zero frequency by performing 
an integral over the remaining time argument. Setting $\tau_0 = n_0 \, 2 \pi/\omega$,
\begin{eqnarray}
S(\Omega=0) \, = \, {\omega \over 2 \pi \,n_0} \int_0^{\tau_0} dt 
\int_{- \infty}^{\infty} dt^{\prime} S(t,t^{\prime}).
\end{eqnarray}
Ref. \onlinecite{MB02} extended the results of Ref. \onlinecite{avron} to the case where this limiting assumption
 is relaxed, yielding a complete description of the quantum statistical properties of an
 adiabatic quantum pump, albeit restricted to small pumping amplitudes. Results
 made use of the 
 generalized emissivity matrix. These results were generalized to arbitrary
 pumping amplitudes by Ref. \onlinecite{Polianski02}. Here, our goal is to address the 
question whether electron-electron interactions affect the pumping noise and how.
 Following formula (14) of Ref. \onlinecite{Polianski02}
 and applying their Eq. (15)  without assuming that the 
 pumping amplitudes $X(t)$ are small,
it is possible to put the zero frequency 
noise $S$ for arbitrary pumping amplitudes into the form
\begin{eqnarray}
S(\Omega=0) \, &=&\,
 {1 \over (2 \pi \hbar)^2}
 {e^2 \over \tau_0} \int_0^{\tau_0}dt \int_{-\infty}^{\infty} d(t^{\prime}-t) \,
\int_{- \infty}^{\infty} f(- \epsilon_1)
\int_{- \infty}^{\infty}  f(\epsilon_2)  \nonumber \\
 & & \, \times \, Tr\Bigl\lbrack
s(\epsilon_1,t)^{\dagger} \sigma_z s(\epsilon_2,t)
s(\epsilon_2, t^{\prime})^{\dagger} \sigma_z
 s(\epsilon_1,t^{\prime})- I \Bigr\rbrack
 e^{i{(t - t^{\prime})(\epsilon_1 - \epsilon_2)\over \hbar}} \, 
d \epsilon_1 d \epsilon_2,
\label{Sbasic}
\end{eqnarray}
where $s(\epsilon_1,t)$ is the $2 \times 2$ $S$-matrix, for an incoming wave at
 energy $\epsilon_1 + \epsilon_F$ and value of the 
pumping parameter $X(t)$.
 $\tau_0$ is a time which is much larger than the period $\tau$.
 $\epsilon_1 = \hbar^2 k_1^2 / 2 m - \epsilon_F$,
 where $k_1$ is a wave vector and $\epsilon_F$ is 
the Fermi energy $\epsilon_F = \hbar^2 k_F^2 / 2m$. 
$f$ is the Fermi-Dirac function. Here, $s(\epsilon_1,t)$ 
is taken for fixed $t$\cite{Polianski02}. 
Since we work at temperature much smaller than $\hbar \omega/k_B$, as explained before, 
we can set 
$f(- \epsilon_1) =1$ for $\epsilon_1\,> \,0$ and $0$ otherwise. 
We set 
$\epsilon^{\prime}_2 = -\epsilon_2$ so
 that both $\epsilon_1$ and $\epsilon^{\prime}_2$ will be positive. 
$M$ is defined as
\begin{eqnarray}
M(\epsilon_1,\epsilon^{\prime}_2,t) \, &=&\, s(\epsilon_1,t)^{\dagger}
 \sigma_z  s(-\epsilon^{\prime}_2,t).
\label{Mequation}
\end{eqnarray}
$S$ now reads
\begin{eqnarray}
S  = 
 {e^2 \over (2 \pi \hbar)^2}
\,  lim_{\tau_0 \rightarrow \infty}
\int_0^{\tau_0} \!\! {dt \over \tau_0} \!
 \int_{- \infty}^{\infty}\!\!\!\!  dt^{\prime} \!
\int_0^{\infty} \!\!\!\!\!  d \epsilon_1 \!   \int_0^{\infty} 
\!\!\!\!\! d \epsilon^{\prime}_2 \,
 Tr \Bigl\lbrace 
M(\epsilon_1,\epsilon^{\prime}_2,t)
 M(\epsilon_1,\epsilon^{\prime}_2,t^{\prime})^{\dagger} \! - I \Bigr\rbrace \,
e^{-i{(t-t^{\prime}) (\epsilon_1 + \epsilon^{\prime}_2) \over \hbar}}.
\end{eqnarray}
Now, for large pumping amplitudes, the above formula needs to be rearranged, 
using the fact that $X(t)$ is a periodic function of
 period $2 \pi / \omega$. 
Note that the dependence of $M$ on $\epsilon_1$ and $\epsilon^{\prime}_2$ prevents
 the direct use of fast Fourier transform. Nevertheless, we can use the fact that,
 for given values of $\epsilon_1$ and $\epsilon^{\prime}_2$, $M(t)$  
and $M(t^{\prime})$ are periodic functions of $t$. 
Switching to Fourier transform
\begin{eqnarray}
{\hat M}_n(\epsilon_1, \epsilon^{\prime}_2) \, &=& \,
{\omega \over 2 \pi} \,\int_{0}^{{2 \pi \over \omega}} 
M(\epsilon_1, \epsilon^{\prime}_2,t) \, e^{-in \omega t} \, dt, \\
M(\epsilon_1,\epsilon^{\prime}_2, \, t) \, &=& \, 
\sum_{n = - \infty}^{+ \infty} {\hat M}_n(\epsilon_1,\epsilon^{\prime}_2) \,
 e^{i n \omega t}.
\end{eqnarray}
Performing the trace, we arrive at
\begin{eqnarray}
S  = {e^2 \over 2 \pi \hbar^2} 
\int_0^{\infty} \!\!\! d \epsilon_1 \!\int_0 ^{\infty} \!\!\! d \epsilon^{\prime}_2
 \!\sum_{n=- \infty}^{\infty}  
\!\Bigl(\vert {\hat M}_{1,1 \, n}\vert^2 + 
\vert {\hat M}_{1,2 \, n}\vert^2 + 
\vert {\hat M}_{2,1 \, n}\vert^2 + 
\vert {\hat M}_{2,2 \, n}\vert^2 - 2 \delta_{n,0}\Bigr) \,
 \delta \Bigl( {\epsilon_1 + \epsilon^{\prime}_2 \over \hbar} - n \omega\Bigr),  
\label{fullformulaforS}
\end{eqnarray}
where
$\delta_{n,0}$ is $1$ if $n=0$ and zero otherwise and
 ${\hat M}_{i,j \, n}$  is the $(i,j)$ element of matrix of
 ${\hat M}_n(\epsilon_1, \epsilon^{\prime}_2)$, where energy dependences
 have been omitted to ease the notations.

When $\hbar \omega$ is much smaller than $\epsilon_F$, formula (\ref{fullformulaforS}) can be
 simplified.
 In this case, $M(\epsilon_1,\epsilon^{\prime} _2,t)$ will be different from
$M(0,0,t)$ only when $\epsilon_1$ or $\epsilon^{\prime}_2$ are a non negligible 
fraction of $\epsilon_F$. This occurs because 
$M(0,0,t)$ corresponds to matrix $M$ for incident wave and outgoing wave
 at energy $\epsilon_F$. We denote by $\epsilon_{1F}$ typical energies of the order
 $\epsilon_F$. $\epsilon_{1F}$  will correspond to
 $n$ of the order $\bigl(\epsilon_{1F}/\hbar \omega\bigr)$, which is very large.
 The Fourier transform
 ${\hat M}_n(\epsilon_1,\epsilon^{\prime} _2)$ will decrease exponentially
 with $n$ for large $n$. Thus, we can neglect the dependence on $\epsilon_1$
 and $\epsilon^{\prime}_2$ and replace them by zero, which amounts to replacing
 the energies by $\epsilon_F$, except in the argument of the $\delta$ function.
 Under these conditions, we have 
\begin{eqnarray}
S \, \simeq \, e^2 {\omega \over \ 2 \pi} 
\Biggl\lbrack
\sum_{n \geq 1}  \, n\, 
\biggl(
\vert {\hat M}_{1,1 \,n} \vert^2 +
\vert {\hat M}_{1,2 \,n} \vert^2 +
\vert {\hat M}_{2,1 \,n} \vert^2 +
\vert {\hat M}_{2,2 \,n} \vert^2
\biggr)
\Biggr\rbrack.
\label{noiseMn}
\end{eqnarray}
With our form of the $S$-matrix, this formula is equivalent to 
Eqs. (24a), (24b) and (24c) of Ref. \onlinecite{MB04}, apart from am overall factor 2. See 
Appendix C for details. 
For numerical simulations however, we did not make this simplification 
 and kept the dependence on $\epsilon_1$ and $\epsilon^{\prime}_2$ 
of Eq. (\ref{fullformulaforS}).
\vspace{2. mm}

\section{Discussion of physical results}

We now illustrate these formulas by computing the charge and noise, in the case of two-delta potentials model. 
The two parameters
 of the drive $X$ and $Y$ (Eq.(9)) are chosen to vary periodically according to: 
\begin{eqnarray}
X \, &=& \, X_0 + \eta \, cos(\omega t), \\
Y \, &=& \, Y_0 + \eta \, cos(\omega t - \varphi),
\end{eqnarray}where $X_0$ is a constant offset potential and $\varphi$ a phase difference. 
 Note that $X$, $X_0$ and $\eta$ are all dimensionless, 
see Eq. (\ref{definitionofVp}). To 
 ensure maximal pumping, we shall specialize 
\cite{GAO05} to $\varphi = \pi / 2$.

\subsection{Zero offset}

\begin{figure}[h] 
\epsfxsize 8. cm  
\centerline{\epsffile{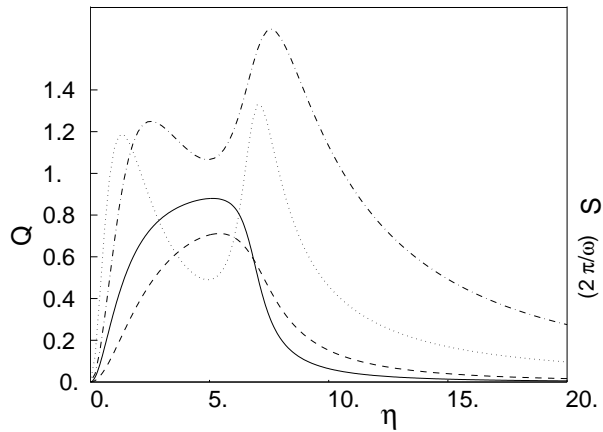}}
\caption{$Q$, charge with interactions, (solid line), 
$Q_0$, charge without interactions, (dashed line), both in units
 of $e$. $S$, noise with interactions, (dotted line) and 
$S_0$ without,  (dashed dotted line), multiplied by 
$2 \pi / \omega$, in units of
 $e^2$, vs. $\eta$. Essential 
parameters are $X_0=0$, no offset,  $l^{2 \alpha} = 0.3$, $k_Fa \, = \, 0.5$ and
 $\hbar \omega = 10^{-2} \epsilon_F$.}
\end{figure} 
\begin{figure}[h] 
\epsfxsize 7. cm  
\centerline{\epsffile{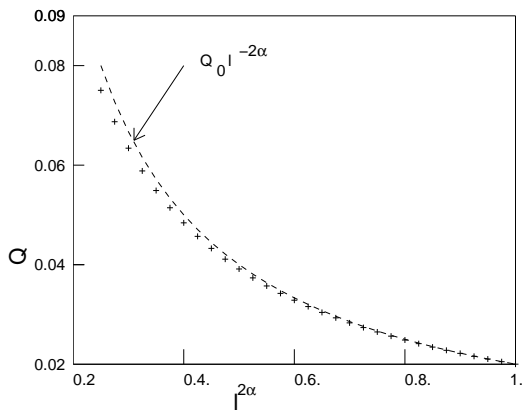}}
\caption{$Q$, charge with interactions, (pluses),
 and $Q_0 \,l^{-2 \alpha}$, (upper dashed line), both in units
 of $e$ vs. $l^{2 \alpha}$, for
 $\eta=0.3$, $X_0=0$, $k_F a \, = \, 0.5$ and $\hbar \omega = 10^{-2} \epsilon_F$.}
\end{figure}

First, the case without offset $X_0=Y_0=0$ is studied. 
To look at the influence of interactions, we plot in Fig. 1
 the pumped charge in units of $e$, with interactions and without,
 versus the amplitude of the drive $\eta$,
 for an interaction parameter $l^{2 \alpha}= 0.3$ 
 (moderate electron-electron interactions). There are three regimes: weak pumping,
 $\eta \ll 1$, intermediate pumping, $\eta$ of order $1$ and large pumping amplitudes, 
$\eta \gg 1$. The current noise times
 $2 \pi / \omega$, in the limit of small $\omega$,
 is plotted together on the same figure in units of $e^2$. 
Analytically, for $\eta \ll 1$, $Q$ reads
\begin{eqnarray}
Q \, =\,{e \over 4} sin(4 k_F a)\, l^{-2 \alpha} \eta^2.
\end{eqnarray}
 
As noted in Ref. \onlinecite{rao}, in the weak pumping regime, 
charge $Q$ is larger with interactions by a factor $l^{-2 \alpha}$, see Fig. 2.
Results of Ref. \onlinecite{MB02} for the noise, 
valid for weak pumping and no interactions can be adapted in a straightforward fashion to
 the case with interactions. We find the following formula for the noise for weak pumping. 
\begin{eqnarray}
S \, =\, e^2 l^{-2 \alpha} \eta^2 {\omega \over 2 \pi}.
\end{eqnarray}

\begin{figure}[h] 
\epsfxsize 7. cm  
\centerline{\epsffile{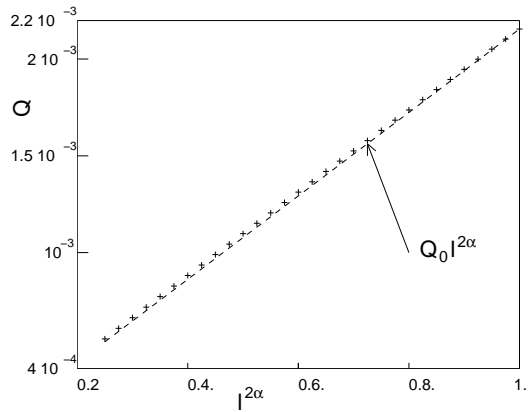}}
\caption{$Q$, charge with interactions, (pluses),
 and $Q_0 \,l^{2 \alpha}$, (lower dashed line), both in units
 of $e$, vs. $l^{2 \alpha}$, for $\eta=15$, $X_0=0$, $k_F a \, = \, 0.5$ and
 $\hbar \omega = 10^{-2} \epsilon_F$. In this regime, $Q$ is approximately larger
 than $Q_0$ by a factor $l^{- 2 \alpha}$.}
\end{figure} 
\begin{figure}[h] 
\epsfxsize 7. cm  
\centerline{\epsffile{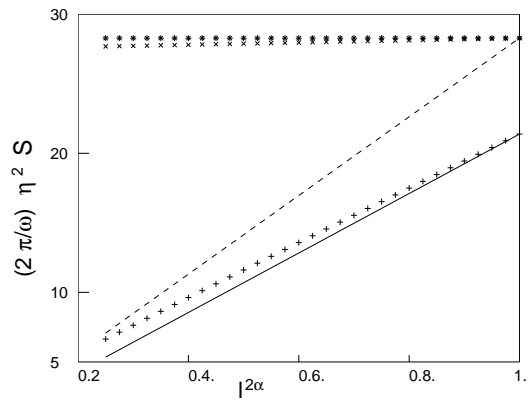}}
\caption{$(2 \pi / \omega) \, \eta^2 S$, noise with interactions scaled by $\eta^2$ and by 
the period, $(+)$
 and $(2 \pi / \omega) \, \eta^2 S_0 l^{2 \alpha}$, lower solid line, for moderate $\eta =15$. 
Also illustrated by the top two curves, is
 the very large pumping amplitude regime, 
$(2 \pi / \omega) \,\eta^2 S$ ($\times$) and 
$(2 \pi / \omega) \, \eta^2 S_0$ ($*$), for very large $\eta$; $\eta=100$.
 The top two curves are close to each other. 
An attempt to fit $\eta^2 S$ by $l^{2 \alpha} \eta^2 S_0$ (dashed line) for $\eta=100$, 
clearly fails for this regime of pumping amplitudes. $k_F a =0.5$ and 
 $\hbar \omega = 10^{-2} \epsilon_F$ for all cases.}
\end{figure} 

The noise is thus increased by the same factor as the 
current. The Fano factor, defined as the ratio $S/e \langle I \rangle$,
 is $4 / sin(4 k_F a)$ and 
remains independent of the interactions, as long as we remain in the 
 weak pumping regime. This corresponds to the very left part of Fig. 1, for
 $\eta$ smaller than $0.25$, typically.

At intermediate pumping amplitudes, $Q$ reaches a maximum value $Q_{max}$ which is 
again larger than its non-interacting analog $Q_{0  \, max}$. This maximum is of 
the order of the single electron charge, but less than it.
 Meanwhile, the noise decreases. This is a reminder
 of the reduction of the noise by a factor $T (1-T)$,
 where $T$ is the electrons transmission coefficient in quantum wires. 
This explains why the noise exhibits a first maximum around 
$\eta$ close to $1$, since $Q$ gets closer to one electron charge, noise will decrease.
 Then, for moderate amplitudes, $\eta$ around $6$, charge decreases and passes through
 the value $0.5 e$, this corresponds then to the second maximum of the noise.

For large, but not very large  pumping amplitudes,
 typically $\eta =10$, $Q$ remains smaller than $Q_0$ but behaves in the same way,
 namely as $\eta^{-3}$, as noted in Ref. \onlinecite{GAO05}. 
As a function of the interaction parameter, 
$Q$ behaves as $Q_0 l^{2 \alpha}$. See Fig. 3.
 For very large pumping amplitudes, (typically order $100$ or more),
 $Q$ becomes practically equal to $Q_0$,
 $Q-Q_0$ behaves as $\eta^{-4}$, see appendix B for details. 

For the noise, we found numerically that $S$ and its analog  without interactions, $S_0$,
both decrease as $\eta^{-2}$, much slower than the charge. See below for analytical derivations.
 As concerns now the interaction dependence of the noise,
 $S$ is always smaller than $S_0$, but for very large $\eta$,
 $S$ tends towards $S_0$. More precisely, for large but still reasonable
 $\eta$, of the order $10$ typically,
 $S$ is almost equal to $S_0 l^{2 \alpha}$, whereas for very large $\eta$,
 of the order $100$ or more, $S$ and $S_0$ are practically the same. This is not surprising
 since $Q$ and $Q_0$ are then also practically equal in the end.

 This dependence on
 the interaction parameter is shown in fig. 4. We have to plot
 $(2 \pi / \omega) \, S \eta^2$, vs. $l^{2 \alpha}$, but the overall factor 
$2 \pi / \omega$ is unimportant; the product $S \eta^2$ 
 can be compared to both $l^{2 \alpha} S_0 \eta^2$ and to $S_0 \eta^2$
 for $\eta=15$ and $\eta=200$. For $\eta=15$, we see that $\eta^2 S$ is fairly well approximated
 by $\eta^2 S_0 l^{2 \alpha}$. On the contrary, for $\eta=200$, such a fit fails and instead,
 $\eta^2 S$ is almost equal to $\eta^2 S_0$.

The results at large $\eta$ 
can be derived from analytical formula for the charge and noise. An expansion 
for large $\eta$ is performed, $X$ and $Y$ behave
 as $\eta$, except at particular points where $X$ or $Y$ are zero. See Appendix A for details.

\subsection{Non-zero offset}

We now turn to the case where $X_0$ is non-zero, which enables to have regions
 where $Q$ is almost quantized. There are basically three cases, according
 to the value of $k_F a$.

 The first case corresponds to $k_F a = n \pi / 2$ (rigorously), where
 $n$ is an integer.
In this case, it is impossible to pump anything. The reason is given below. 
The second case corresponds to the case where $k_F a$ is small but non-zero,
 $0.1$ typically. We first describe the behavior, then give numerical illustrations and last
 provide analytical justifications.  In this case, the charge is almost zero up to $\eta=X_0$.
 It rises quickly around $\eta = \sqrt{2} X_0$ and reaches a value close
 to quantized $e$ for a wide range of values of $\eta$. This is the quantized 
region of $\eta$. The width of this region can be shown
 to scale approximately as $(k_F a)^{-1}$. After the end of this region, 
$Q$ and $Q_0$ first decrease abruptly and for even larger values 
of $\eta$, decrease slower, as $\eta^{-3}$. The noise in the quantized
 region and around it seems to be well approximated by $Q (e-Q)$, reminiscent
 of the noise for fermions in narrow quantum wires.
 However, this does not last when $\eta$ becomes noticeably 
out of the region of almost quantized charge, 
since $Q$ and $Q_0$ behave as $\eta^{-3}$, whereas
 $S$ and $S_0$ decay only as $\eta^{-2}$.
 
Fig. 5 shows noise and charge with and without interactions, versus $\eta$ for $k_Fa =0.1$. 
Fig. 6 shows a comparison between $S$ and a least square fit of the form 
$C \, Q(e-Q)$, in the quantized charge regime, where $C$ is the only adjustable parameter. 
See captions for details. 
\begin{figure}[h] 
\epsfxsize 8. cm  
\centerline{\epsffile{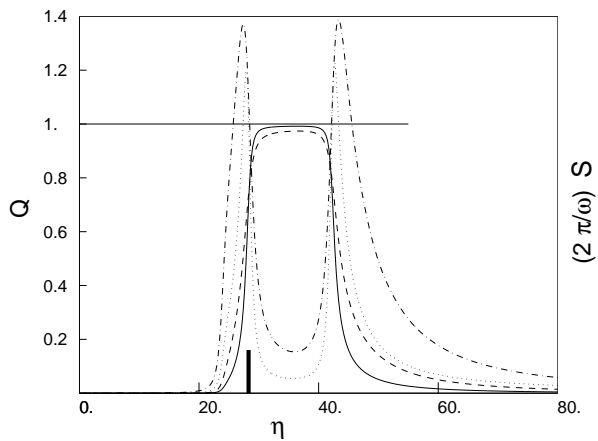}}
\caption{$Q$, (solid line), curve with a plateau for $\eta$ between $28$ and $43$, (the 
 quantized region), $Q_0$, (dashed line). The two curves with
 two spikes located at $\eta$ around $28$ and at $\eta$ around $43$ represent 
$(2 \pi / \omega)  \, S$, (dotted line) and
 $(2 \pi / \omega) \, S_0$, (dashed dotted line),  vs. $\eta$. Essential 
parameters are $X_0=20$, $l^{2 \alpha} = 0.3$, $k_Fa \, = \, 0.1$ and 
 $\hbar \omega = 10^{-2} \epsilon_F$. Electron charge $e$ is set to $1$.
 The solid horizontal line of ordinate $1$ and the
 thick vertical line at $\eta = X_0 \sqrt{2} \simeq 28.28$ 
are guides to the eye.}
\end{figure} 
\begin{figure}[h]
\epsfxsize 6. cm  
\centerline{\epsffile{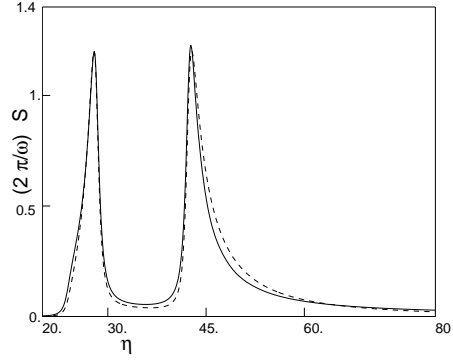}}
\caption{$S$ (solid line), in units of $e^2$, and best fit of the form $C\, Q(e-Q)$
(dashed line) in the interval $20 \leq\eta\leq 45$, for
 the same situation as Fig. 5. The fit no longer works in the large pumping regime, 
for $\eta$ between $45$ and $80$ here.}
\end{figure}
\vspace{ 4. mm}
We now turn to analytical justifications of the previous assertions. 

We now explain
 why pumping is impossible for $k_Fa = n \pi / 2$. 
Since $sin(2 k_F a)=0$, 
the scattering matrix now depends only on a single parameter, 
the combination $({\overline X}+ {\overline Y})$, 
see eqs. (\ref{equationsforsmatrix1}-\ref{equationsforsmatrixlast}), so we denote
 by $s_{ij}^{(0) \, \prime}$ its derivative with respect to ${\overline X} + {\overline Y}$.
Thus,
\begin{eqnarray}
Q_0 \, =\, {e \over \pi} \int_0^{{2 \pi \over \omega}} 
Im  \biggl\lbrack \Bigl(\sum_{j=1}^2 s_{1j}^{(0) \, \prime}({\overline X} + {\overline Y}) 
\, s_{1j}^{(0) \,*}({\overline X} + {\overline Y})  \Bigr) 
\biggr\rbrack
\, {d \over dt}({\overline X} + {\overline Y}) \, \, dt \, = \, 0,
\end{eqnarray}
because the bracket is just
 $(1/2) \, d/d({\overline X} + {\overline Y})\,
 \bigl(\vert s_{11}^{(0)}\vert^2  + \vert s_{12}^{(0)}\vert^2 \bigr)$. 
Then, we look at the case when $k_F a$ is close to $n \pi / 2$, but different from it. 
Clearly, when $sin(2k_F a)$ is small, for
 $\eta \vert sin(k_F a) \vert < 1$, we will be back to the former case, so
 $Q_0$ can start to level noticeably from $0$ only for $\eta$ values larger than a critical value
 $\eta_{c1}$, which is proportionnal to $1/\vert sin(2 k_F a)\vert$, independently of $X_0$.
 There is at least another scale, namely $X_0$. For $X_0$ large,
 (typically larger than $10$)
 and $\eta$ smaller than $X_0$, the pumping contour is 
a circle which does not enclose the origin and both transmission
 $s_{12}^{(0)}$ and derivatives of the transmission coefficients,
 $\partial s_{12}^{(0)} /\partial X$ are small. $Q_0$ will remain very small.
Thus, to have
 a significant $Q_0$, one needs $\eta > max \bigl(X_0, 1/\vert sin(2 k_Fa)\vert\bigr)$,
 where $max(x,y)$ is the larger of $x$ and $y$. For even larger $\eta$, 
when terms like $sin(2 k_F a) {\overline X} \, {\overline Y}$
 dominate over terms linear in ${\overline X}$ or ${\overline Y}$, i.e.
 $\bigl(\vert sin(2 k_F a)\vert \, \eta\bigr) \, \gg 1$,
 it is possible to expand in $\eta^{-1}$ and
 we are back to the large pumping regime where $Q$ decays as $\eta^{-3}$.
So, for $\vert sin(2 k_Fa) \vert$ smaller than $X_0$, there will be a region between
 $max(X_0, C_1/sin(2 k_F a))$ and $C_2 /sin(2 k_F a)$, where
 $C_1$ and $C_2$ are constants, where $Q_0$ is appreciable. These
 very qualitative arguments however do not
 prove that the charge is almost quantized in this interval,
 whose width is  of the order $\vert sin(2 k_Fa)\vert^{-1}$.

We now turn to the analytical explanation of the $Q (e-Q)$ behaviour in the quantized charge 
regime. It is useful to disentangle the effects of the fluctuations of $T$ and
 those of the phase $\theta$.
 We start from (\ref{noiseMn})  and use the fact that for any periodic function
 $f(x)$, if ${\hat f}_n$ denotes the Fourier component at frequency $n \omega$,
 denoting $x = \omega t$,
\begin{eqnarray}
\sum_{n \geq 1} n \vert {\hat f}_n \vert^2 \, =\, 
\int_0^{2 \pi} \!\int_0^{2 \pi} 
{\vert f(x) - f(x^{\prime}) \vert \over (x-x^{\prime})^2} {dx \, dx^{\prime}  \over 4 \pi^2}.
\end{eqnarray}
Applying the former equality to ${\hat M}_{1,1 \, n}$,
 ${\hat M}_{1,2 \, n}$,
 ${\hat M}_{2,1 \, n}$
 and ${\hat M}_{2,2 \, n}$, and denoting 
by $M_{11}(x)$ the inverse Fourier transform of ${\hat M}_{1,1 \, n}$,
we get
\begin{eqnarray}
S  \, &=& \, e^2 {\omega \over 2 \pi} \! \int_0^{2 \pi} \!\! \int_0^{2 \pi} 
\! \biggl(
\vert M_{11}(x) - M_{11}(x^{\prime}) \vert^2 +
\vert M_{12}(x) - M_{12}(x^{\prime}) \vert^2 +
\vert M_{21}(x) - M_{12}(x^{\prime}) \vert^2 +
\vert M_{22}(x) - M_{22}(x^{\prime})\vert^2 \biggr)\nonumber \\
&\,\,&\,\,\,\,\,\,\,\,\,\,\,\,\,\,\,\,\,\,\,\,\,\,\,\,\,\,\,\,\,\,\,\,\,\,\,\,\, 
(x-x^{\prime})^{-2} \,\,{dx \, dx^{\prime} \over 4 \pi^2}. 
\end{eqnarray}

In terms of $\theta$ and $T$, using (\ref{sm}) and (\ref{Mequation}), this reads
\begin{eqnarray}
S \, = \, 8 e^2 \Bigl({\omega \over 2 \pi}\Bigr) \Bigl(I_1 + I_2\Bigr),
\end{eqnarray}
with
\begin{eqnarray}
I_1 \, &=& \, \int_0^{2 \pi} \! \int_0^{2 \pi} 
{\vert T(x) - T(x^{\prime})\vert^2 \over
 (x-x^{\prime} )^2} \, {dx \, dx^{\prime} \over 4 \pi^2}, 
\label{I1}\\
I_2 \, &=& \, \int_0^{2 \pi} \! \int_0^{2 \pi} 
{\vert g(x) - g(x^{\prime})\vert^2 \over
 (x-x^{\prime})^2} \, {dx \, dx^{\prime} \over 4 \pi^2},
\label{I2}
\end{eqnarray}
with $g(x) = \sqrt{T(1-T)} \, e^{i \theta}$.
We expand $\vert g(x) -g(x^{\prime})\vert^2$ and rewrite $I_2$ in the form
\begin{eqnarray}
I_2 \, =\, J_1 + J_2,
\end{eqnarray}
with
\begin{eqnarray}
J_1 \, &=& \,  \int_0^{2 \pi}  \!  \int_0^{2 \pi}  
\Biggl\lbrack
{ \sqrt{T(x) \bigl(1-T(x)\bigr)} -  \sqrt{T(x^{\prime}) \bigl(1-T(x^{\prime})\bigr)}
 \over (x-x^{\prime})}\Biggr\rbrack^2 {dx  \, dx^{\prime} \over 4 \pi^2}, \\
J_2 \, &=& \,
 \int_0^{2 \pi}  \!  \int_0^{2 \pi} 
\sqrt{T(x) \bigl(1-T(x)\bigr)}
\sqrt{T(x^{\prime}) \bigl(1-T(x^{\prime})\bigr)}
{\vert e^{- i \theta(x)} - e^{-i \theta(x^{\prime})}\vert^2 \over (x-x^{\prime})^2} \, 
{dx  \, dx^{\prime} \over 4 \pi^2}.
\end{eqnarray}
Thus, noise breaks into three parts, one related to the fluctuations of $T$,
 another to the fluctuations of $\sqrt{T (1-T)}$, and the last
 to the fluctuations of $\theta$, in fact, to the variations of
 the slope $d\theta/dx$, since $\theta$ has to vary by $2 \pi$ in one cycle 
to get appreciable pumped charge. The physical message is that if the fluctuations 
of $T$ are much smaller than the fluctuations of the phase, then noise
 shows a $Q (e-Q)$ behaviour. Otherwise, fluctuations of $T$ bring an extra noise that
 does not contribute to the pumped charge and overall noise is thus larger
 than $Q (e-Q)$. We now try to establish this more firmly.  

\vspace{3. mm}

In this paragraph, we now show that, for $Q_0 \, \leq \, e/2$, 
\begin{eqnarray}
S_0 \, \geq \, 8 \Bigl({\omega \over 2 \pi}\Bigr) \, Q_0 \, (e-Q_0).
\label{S0bound}
\end{eqnarray}
The first integral $J_1$ involves solely the fluctuations of $T$ and can be rewritten as
\begin{eqnarray}
J_1 \, =\,
\sum_{n=1}^{\infty} n \vert  \widehat{ \sqrt{T (1-T)}}_{n} \vert^2.
\end{eqnarray}
We have a lower bound for $J_1$ by replacing $n$ by $1$ in all terms of the sum except the term for $n=0$,
\begin{eqnarray}
J_1 \, \geq \, \sum_{n=0}^{\infty}  \vert  \widehat{ \sqrt{T (1-T)}}_{n} \vert^2 -
 \vert  \widehat{ \sqrt{T (1-T)}}_{0} \vert^2.
\end{eqnarray}
Using then the Parseval identity for the first term and using the notation 
$\langle f \rangle \, \equiv \, \int_0^{2 \pi} f(x) \, dx/2 \pi$ for the second,
\begin{eqnarray}
J_1 \geq 
\Bigl\langle \bigl\vert \sqrt{T(1-T)} \bigr\vert^2 \Bigr\rangle - 
\Bigl\langle \sqrt{T(1-T)} \Bigr\rangle^2 \, = \, 
\langle T \rangle - \bigl\langle T^2 \bigr\rangle - 
\Bigl\langle \sqrt{T(1-T)} \Bigr\rangle^2. 
\end{eqnarray}
Now for $J_2$, applying twice the H{\"o}lder inequality,
\begin{eqnarray}
J_2 \, &=& \, 
\int_0^{2 \pi} {dx^{\prime} \over 2 \pi} 
\sqrt{T(x^{\prime}) \bigl(1-T(x^{\prime})\bigr)}  
 \int_0^{2 \pi} 
\sqrt{ T(x) \bigl(1-T(x)\bigr)} 
{\vert e^{- i \theta(x)} - e^{- i \theta(x^{\prime})} \vert^2 \over (x-x^{\prime})^2}
 \, {d x \over 2 \pi} \nonumber \\
& \geq & \,\int_0^{2 \pi} {dx^{\prime} \over 2 \pi} \,
\sqrt{T(x^{\prime}) \bigl(1-T(x^{\prime})\bigr)}  
\, \Bigl\langle \sqrt{T(1-T)} \Bigr\rangle  
\int_0^{2 \pi} 
{\vert e^{- i \theta(x)} - e^{- i \theta(x^{\prime})} \vert^2 \over (x-x^{\prime})^2}
 {d x \over 2 \pi} \, \nonumber \\
& \geq &
\Bigl\langle \sqrt{T(1-T)} \Bigr\rangle^2 
\int_0^{2 \pi} {dx \over 2 \pi} \int_0^{2 \pi} {dx^{\prime} \over 2 \pi} 
 {\vert e^{- i \theta(x)} - e^{- i \theta(x^{\prime})} \vert^2 \over (x-x^{\prime})^2} .
\end{eqnarray}
For the last double integral, we proceed as before, going again to Fourier transform,
 isolating the component of order $0$ 
and using the Parseval identity, it is larger than
$\bigl\langle \vert e^{i \theta}  \vert^2 \bigr\rangle 
- \vert \langle e^{i \theta} \rangle\vert^2$.
Always, $\vert e^{i \theta} \vert=1$ and, for
 reasonable $\theta(x)$, we can assume symmetry $x$ into $-x$ which implies 
$\langle sin \theta \rangle =0$. We can also assume symmetry when
 $x$ is changed into $\pi -x$, which implies $\langle cos \theta \rangle =0$.
We have
\begin{eqnarray}
J_2 \geq 
\Bigl\langle \sqrt{T(1-T)} \Bigr\rangle^2.
\label{J2}
\end{eqnarray}
Now, for $I_1$,  using the same technique, (going to Fourier
 transform and isolating the $n=0$ component)
\begin{eqnarray}
I_1 \, \geq  \, \langle T^2  \rangle - \langle T \rangle^2.
\end{eqnarray}
Putting everything together
\begin{eqnarray}
I_1 + I_2 \geq
\langle T^2 \rangle - \langle T \rangle^2 + 
\langle T \rangle - \langle T^2 \rangle - \Bigl\langle \sqrt{T(1-T)} \Bigr\rangle^2
 + \Bigl\langle \sqrt{T(1-T)} \Bigr\rangle^2 \, = \, 
\langle T \rangle - \langle T \rangle^2.
\end{eqnarray}

Now, assuming that $\int_0^{2 \pi} {d \phi \over dx} \, {dx \over 2 \pi} =0$,
 (the circulation of $\phi$ is zero in one cycle), 
\begin{eqnarray}
Q \, =\, e \, \Bigl(1 - \Bigl\langle T {d\theta \over dx} \Bigr\rangle \Bigr),
\end{eqnarray}
which implies
\begin{eqnarray}
Q (e-Q) \, =\, e^2 \, \Bigl( \Bigl\langle T {d \theta \over dx} \Bigr\rangle - 
\Bigl\langle T {d \theta \over dx} \Bigr\rangle^2 \Bigr).
\end{eqnarray}
Moreover, by the H\"older inequality
\begin{eqnarray}
\Bigl\langle T {d \theta \over dx} \Bigr\rangle \,
 \geq \, \Bigl\langle T \Bigr\rangle \, 
\Bigl\langle {d \theta \over dx} \Bigr\rangle \, =\, \langle T \rangle.
\end{eqnarray}
Now, for any $y \geq 1/2$, $y(1-y)$  is a decreasing function of $y$.
Thus, if $\Bigl\langle T {d \theta \over dx} \Bigr\rangle \, \geq \, 1/2$ , i.e.
 $Q \, \leq \, e/2$, then
\begin{eqnarray}
\langle T \rangle - \langle T \rangle^2 \, \geq 
\Bigl\langle T {d\theta \over dx}  \Bigr\rangle - 
\Bigl\langle T {d \theta \over dx} \Bigr\rangle^2.
\end{eqnarray} 
Thus $S \, \geq 8 \bigl({\omega \over 2 \pi}\bigr) Q_0 (e-Q_0)$, which is (\ref{S0bound}).

\vspace{4. mm}

Now we turn to the case of interest, when $Q_0 \, \geq \,  e/2$, for example in
 the quantized region. We were not able to provide a general 
analytical proof of $Q_0 (e-Q_0)$ behaviour. We first look at simple limiting cases.
 Then, we examine the particular case of the two delta-potentials model.
 
First, in a situtation where $T(x)$ is a constant $T$, (with $T$ small to have almost
 charge quantization), then $Q_0 = e (1-T)$ and
 $S_0 \,=\, 8 C_0 \bigl({\omega \over 2 \pi} \bigr) Q_0 \, (e-Q_0)$ with
$C_0$ depending on the shape of $\theta(x)$, but always $C_0$ is greater than $1$.
 In the case of constant slope $d \theta/dx \, = \, 1$, $C_0=1$.
Second, in the situation where $d\theta/dx$  is constant but $T(x)$ arbitrary,
 we have
 $\langle T \, ( d \theta/dx ) \rangle \, =\, \langle T \rangle$
 and thus
again (\ref{S0bound}) holds for any $Q_0$, not just for $Q_0$ smaller than $e/2$. 

\vspace{3. mm}

However, in practise, $\theta$ varies abruptly by $\pi$ in the vicinity of resonances. This is 
different from optimal pumping strategies which have been studied before\cite{AndreevKamenev,avron}.
We first give qualitative arguments and then give precise calculations for
the model studied here. Let us look at 
the contribution of $J_2$, Eq. (\ref{J2}), to the noise.
 When $x$ and $x^{\prime}$ are both close to resonances
 $\vert e^{- i \theta(x)} - e^{-i \theta(x^{\prime})} \vert^2 /(x-x^{\prime})^2$
 behaves as $(d \theta / dx)^2$ for $(x-x^{\prime}) (d\theta/dx) <1 $.
 If meanwhile, $T(x)$ does not vary too much and assumes the value $T_i$,
 the contribution of this region in the plane $(x,x^{\prime})$ to $J_2$ will be $T_i (1-T_i)/4$. 
The $1/4$ comes from the fact that $\theta$ varies suddenly only by $\pi$ and not by $2 \pi$ 
at each resonance. 
Another contribution will come from the regions where $x$ 
is within the resonance but $x^{\prime}$ just outside or vice versa. 
Let us take $x^{\prime}$ outside to be precise.
Then $\vert e^{-i \theta (x)} - e^{- i \theta(x^{\prime})}\vert = 2$ and integration 
 on $x$ and $x^{\prime}$ will give a contribution mainly from 
$x^{\prime}$ just outside, due to the rapidly decreasing 
factor $(x-x^{\prime})^{-2}$. This will eventually 
 give another factor $T_i (1-T_i)/4$. Despite the non-local character of the integrand in 
$J_2$, regions where $x$ and $x^{\prime}$ are both
 far from a resonance make very little
 contribution to $J_2$. For $Q$, one gets $Q= e(1 - \sum_i T_i/2)$.
In the simple case where there are only two resonances and the $T_i$'s are equal,
 then $Q = e (1 - T_1)$ and $J_2 = T_1 (1-T_1)$. If $I_1$ and $J_1$, which are
 related to the fluctuations of $T(x)$ are much smaller than $J_2$, then, this
 leads to $S_0 = 8 \, (\omega / 2 \pi) \, Q_0 (e-Q_0)$. 

\vspace{3. mm}
We now test the former very qualitative ideas by analytical calculations 
on our particular model. A first thing to be noted is that 
the phase $\phi$ does not intervene in the noise, which is normal since
 noise is related to the two-particle scattering matrix.
 On the contrary, it does formally enter the equation for the pumped charge, 
see Eqs. (\ref{sm}),(\ref{Q0sint}) and (\ref{Qsint}). Nevertheless, the variation of $\phi$ when
 $\omega t$ varies in one period, is always zero so that $\phi$ does not
 play any role. 
This is due to the fact that $\phi$ is the phase of $D$,
 see Eqs. (\ref{s12}) and (\ref{equationsforsmatrixlast}).
 If the real part of $D$ becomes negative, then, its imaginary part cannot be zero
 and thus, $\phi$ can never be equal to $-\pi$ or $\pi$. Moerover, $\phi$ varies 
by strictly less than $2 \pi$ during one period and thus, its circulation is zero, giving no 
 contribution to $Q$ and $Q_0$. 

We now turn to the variations of $\theta$.
 Charge quantization necessitates $\eta$ larger than $\sqrt{2} \, X_0$ and
 $k_Fa$ small. We thus approximate $cos(2 k_Fa)$ by $1$ and set
 $u = sin(2 k_F a) \, \ll \, 1$.
$\theta$ can be written as
\begin{eqnarray}
\theta \, =\, arg(n),
\end{eqnarray}
with
\begin{eqnarray}
n \, =\, {\overline X}\, {\overline Y}\, u \, + {\overline X} + {\overline Y} +
 i u ({\overline Y} -  {\overline X}).
\end{eqnarray}
There exist two values of $\eta$, $\eta_1$ and $\eta_2$, which play a particular
 role.
 For $\eta_1 \leq \eta\leq \eta_2$, when $\omega t$ varies by $2 \pi$, $Re(n)$ changes 
twice its sign and $\theta$ varies by $2 \pi$. In the model studied here,
$\eta_1= X_0 \sqrt{2}$ and $\eta_2 = X_0 \sqrt{2} + 2 \sqrt{2} u^{-1}$.
 Outside this interval of $\eta$, the increase of $\theta$ when $\omega t$ varies
 by $2 \pi$ is zero, not $2 \pi$.
The reason is the following.
For $\eta \leq \eta_1$, $Re(n)$ is always positive and the phase
 $\theta$ remains confined in an interval contained in
 $\lbrack - {\pi \over 2}  \, , \, {\pi \over 2} \rbrack$.  
 For $\eta > \eta_2$, $Re(n)$ changes four times its sign but the contour described by
 $n$, in the complex plane, as $\omega t$ is varies by $ 2 \pi$,
 does not enclose the origin. It can be seen directly, for it is impossible to have
 $Im(n)=0$ and $Re(n) \leq 0$ at the same time. Thus, it is not surprising
 that the quantized region corresponds approximately to the interval 
$\lbrack \eta_1, \, , \, \eta_2 \rbrack$. 

We now examine more precisely the variations of $\theta$ and $T$. 
Apart from a small variation around $x \equiv \omega t = - {3 \pi/4}$,
 $\theta$, as a function of $x$, is essentially flat, except around
 two points $x_1$ and $x_2$. In practise, $x_1$ is close to
 $- {\pi \over 4}$ and $x_2$ close to $3 \pi /4$.
Around those two points, $\theta$ increases fastly by almost $\pi$ each time.
 We can thus model the function $\theta$ by
\begin{eqnarray}
\theta \, &=& \, - \pi, \,\,\,\, {\rm for} \,\,\,\,\,\,\,\,\,\, x \leq x_1-l, \nonumber \\
\theta \, &=& \,  {\pi \over 2} \Bigl({x - x_1 \over l} -1\Bigr) 
\,\,\,\, {\rm for} \,\,\, x_1- l \leq x \leq x_1+l, \nonumber \\
\theta \, &=& \, 0, \,\,\,\,\,\,\,\,\,\,
\,\,\,\, {\rm for} \,\,\, x_1+l \leq x \leq x_2-l, \nonumber \\
\theta \, &=& \,  {\pi \over 2} \Bigl(1 + {x-x_2 \over l}\Bigr)
\,\,\,\, {\rm for} \, \leq x_2-l \leq x \leq x_2 +l, \nonumber \\
\theta \, &=& \,  \pi, \,\,\,\,\,\,\,\,\,\,\,{\rm for} \,\,\, x \geq x_2+l,
\label{thetaparticular}
\end{eqnarray}
$l$ being a small distance.
As for $T$, $\,T(x)$ shows a large peak around $x=x_s = - 3 \pi / 4$, and
 two smaller peaks, practically identical, centered around $x_1$ and $x_2$.
 Away from these values of $x$, $T(x)$ is 
almost zero.

Now, we look at the implications for the charge and noise. When calculating
 the pumped charge without interaction via the formula 
${1 \over 2 \pi} \int_0^{2 \pi} {d \theta \over dx} \, \bigl(1-T(x)\bigr) \, dx$,
 the region around $x=x_s$ does not bring much contribution because variations of
 $\theta$ are small here. 
For the calculation of $Q_0$ and $S_0$, 
we can ignore the large peak in $T$ and thus model $T(x)$ by
\begin{eqnarray}
T(x) \, &=& \, T_1 \, 
\exp \biggl({ (x-x_i)^2  \over (x-x_i)^2-l^2}\biggr)
 \,\,{\rm for} \,\,\, \vert x - x_i \vert \leq \, l, \nonumber \\
T(x) \, &=& \, 0 \,\,\, {\rm otherwise}.
\label{Tparticular}
\end{eqnarray}
$T$ has to be derivable in order to avoid
 logarithmic divergencies due to the factor 
$1/(x-x^{\prime})^2$ in Eqs. (\ref{I1}) and (\ref{I2}).
 Then, it is possible to perform analytical calculations which give 
\begin{eqnarray}
Q_0 \,&=& \, e \bigl(1 - C_1 T_1 \bigr), \\
S_0 \, &=& \, C_2   \Bigl({\omega \over 2 \pi} \Bigr) 
\, (e-Q_0) \, \bigl\lbrack e (1- C_3) + C_3 Q_0) \bigr\rbrack.
\end{eqnarray}
Detailed calculations, involving the integrals $I_1$ and $I_2$ and the constants
 $C_1$, $C_2$ and $C_3$ are given in the appendix.
$C_3$ is smaller than  $1$, (approximately $0.58$).  Even if it is not exactly of the form
 $Q_0 (e-Q_0)$, when there is good charge quantization,
 i.e. when $Q_0$ is not far from $1$, $S_0$ goes as $(e-Q_0)$.
 The essential thing is that $T$ and $\theta$ vary rapidly around certain
 values of $\omega t$. We do not get exactly $Q_0(e-Q_0)$ because 
$T$ varies substantially when $\theta$ jumps.  
One might wonder if the results here are particular to our model. In fact, 
brisk variations of the phase are widely shared by many types of models\cite{EntinWohlman,Levinson}.

\vspace{2. mm}

The third case corresponds to $k_F a$ not close to $n \pi / 2$.
In this case, for large $X_0$, $Q$ is almost zero except in the vicinity 
of a value $\eta_c$, which is very near $\sqrt{2} X_0$; numerically it seems 
 that $\eta_c$ is always a little less than this value.
 The maximum pumped charge is
 of order $e$ but no longer close to one electron charge. Noise has a 
 double peak structure around  $\eta_c$. 
A rough qualitative picture of this can be seen in Eq. (\ref{s12}), 
because, as soon as the integration contour does not get close
 to the point $X(t) =Y(t) =0$, for any $t$, the integrands in Eqs. (\ref{Q0sint}) and
(\ref{Qsint}) are very small. An example is shown in Fig 7.

\begin{figure}[h] 
\epsfxsize 7. cm  
\centerline{\epsffile{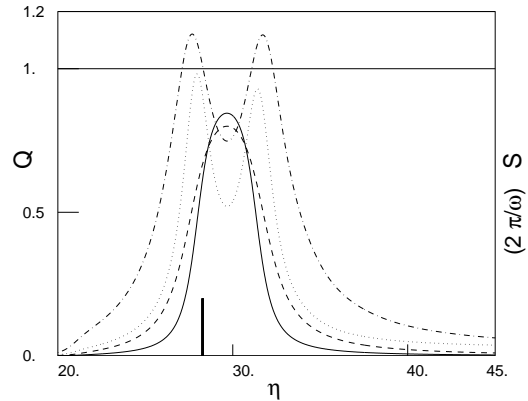}}
\caption{Same as Fig. 5, except that $k_Fa \, = \, 0.4$. The solid 
horizontal line
 of ordinate $1$ and the thick vertical line
 at abscissa $X_0 \sqrt{2}$ are just
 guides to the eye.}
\end{figure} 
\section{Conclusion}
\label{sec:conclusion}

We have studied the influence of weak electron-electron interactions on pumped charge and noise
 in the adiabatic regime, in a mesoscopic one-dimensional disordered wire. Within
 the two-delta potentials barrier model, analytical results were 
obtained for the charge and noise. Results were analyzed numerically for local pumping fields with
 a harmonic dependence. Without any voltage offset, at weak pumping amplitudes, interactions tend to 
enhance the pumped charge, as $l^{-2 \alpha}$, where $l$ is the interaction parameter.
 For fairly large pumping amplitudes, it is exactly the reverse, 
$Q$ and $Q_0$ both decrease as $\eta^{-3}$, 
 but $Q$ remains smaller than $Q_0$ by a factor $l^{2 \alpha}$. At very 
large pumping amplitudes, $Q$ and $Q_0$ are practically the same. As to the pumping noise,
 at weak amplitudes, it increases with interactions, but in the same way as the charge, so that,
 the Fano factor remains constant, independent of the interactions. For moderate pumping amplitudes,
 noise has a double peak structure around the maximum of pumped charge. For large amplitudes, the noise
 decreases slower than the charge, as $\eta^{-2}$, and for very large amplitudes, noise with and without
 interactions become approximately the same. 

As emphasised in Ref. \onlinecite{rao}, interactions
 tend to make resonances sharper, which is conducive to 
obtaining an almost quantized  pumped charge. However, it is not sufficient to enclose a resonance, it is 
 also necessary that the pumping contour does not go too far from the resonance. Otherwise,
 the noise is appreciable and the signal $Q$ can even be very small.

In the case of constant offset $X_0$, the behavior depends if we are close to a resonance,
 $k_F a = n \pi /2$ in the two-delta potentials model. 
  Close to a resonance, there is a region of almost quantized pumped charge where the noise
 seems to follow a $Q(e-Q)$ behavior, reminiscent of the noise reduction
 in quantum wires for good transmission by a $T(1-T)$ factor, where $T$ is the modulus
 of the energy transmision coefficient. Quite generally, noise breaks up
 into pieces due to fluctuations of $T$ and those due to fluctuations of $\theta$.
 We believe that in the quantized region, the fluctuations of $\theta$ are predominant
 and give rise to the $Q(e-Q)$ behaviour.

Interactions do help in having
 a charge closer to $e$ and to reduce the noise. However, it does 
not change the range of pumping amplitudes, where quantized charge is observed,
 i.e. the width of the quantized region practically 
does not depend on the interactions. Qualitative arguments seem to indicate that this
 width scales as $sin(2k_F a)^{-1}$ close to $k_F a = n \pi / 2$.
 Further from the resonance, the maximum charge which 
 can be pumped becomes of order, but less than $e$. Moreover, the region of quantized
 charge shrinks to a narrow window of pumping amplitudes around a value close to $X_0 \sqrt{2}$.   

In summary, our study of noise shows that interactions tend to increase the
 quality of pumping. However, two 
conditions need to be met; first, to operate at certain wavevectors favouring
 sharp resonances and second to have a 
pumping contour which encircles the resonant point, passing  not very far from it. 
Otherwise, only noise is produced and the quantization of the charge is not achieved. These
 restrictions were not pointed out in previous works. In addition, in the quantized
 charge region, noise vanishes as $e-Q$.

\acknowledgments
One of us (T.M) acknowledges the support from  ANR grant ``molspintronics'' of the
 French ministry of research. V.G. acknowledges the kind hospitality of the 
Centre de Physique Th\'eorique de Marseille, where inital parts
 of this work were done.

\appendix
\section{}

In this appendix, we consider the limit of large pumping. 
In order to explain  why, for offset $X_0=0$,  $Q_0$ behaves like $\eta^{-3}$ at
large $\eta$,
 we use Eqs. (\ref{Q0sint}),
(\ref{equationsforsmatrix1}),
(\ref{s12}) and
(\ref{equationsforsmatrixlast}).
 Let us look first at the terms involving 
 $d {\overline X} / dt$. 

${\overline X}$, ${\overline Y}$, as well as their time derivatives
 will be of order $\eta$, except at isolated particular points.
The term $\Bigl\vert \partial \ln s_{12}^{(0)} /\partial {\overline X}\Bigr\vert^2$
 can be expanded in powers
 of $1/{\overline X}$ and $1/{\overline Y}$:
\begin{eqnarray}
\Biggl\vert {\partial \ln s_{12}^{(0)} \over \partial {\overline X}}\Biggr\vert^{\, 2} \, =\,
{1 \over {\overline X}^{\, 2}} \, - \,
 {cotan(2 k_F a) \over {\overline X}^{\, 3}}
 + {3 \, cotan^2(2 k_F a)-1 \over 4{\overline X}^{\, 4}} 
 - {1 + cotan^2(2 k_F a) \over 4 {\overline X}^{\, 2} \, {\overline Y}^{\, 2}} \, + \,
{cotan^2(2 k_F a)-1 \over 2 {\overline X}^{\, 3} \, {\overline Y}} + o(\eta^{-5}).
\end{eqnarray}
All terms multiplied by $d{\overline X}/dt$ and then,
 integrated over one period
 give zero. Note that a term like $1 /({\overline X}^{\, 4} \,{\overline Y})$,
 which is in $\eta^{-5}$, would not give $0$.  
Thus, for large $\eta$, the term proportionnal 
to $d {\overline X} / dt$ in $Q$, behaves at least as $\eta^{-3}$.

For the term  involving $d\overline Y / dt$, 
the situation is simpler. $d{\overline Y} / dt$ goes as $\eta$,
 but since $D$ goes as $\eta^2$, $T_0$ goes as $\eta^{-4}$,
 thus this term is at least in $\eta^{-3}$. 

Now the remainder of contributions to $Q-Q_0$ behave at least  like $\eta^{-4}$ for large
 $\eta$; it can be seen from (\ref{Qsint}). 
For large $\eta$, since $T_0$ goes as $\eta^{-4}$, 
the quantity $T_0/ (1 + T_0(A^{2 \alpha}-1))$
 is practically equivalent to $T_0 \sim \eta^{-4}$. 
 $Im \lbrace \partial \ln s_{12}^{(0)} / \partial {\overline X} \rbrace$ goes as
 $\eta^{-1}$ for large $\eta$. The same holds for 
$Im \lbrace \partial \ln s_{12}^{(0)}/ \partial {\overline Y} \rbrace$. 
In the integral in the r.h.s. of (\ref{Qsint}),
$Im \lbrace \partial \ln s_{12}^{(0)}/ \partial {\overline X} \rbrace$ is
 larger, by $\eta$, than 
$\vert (\partial \ln s_{12}^{(0)} / \partial {\overline X})\vert^2$. 
Then,   
$Im \lbrace \partial \ln s_{12}^{(0)} / \partial {\overline X} \rbrace$ 
 is of order $\eta^{-1}$ whereas $T_0$ is of order $\eta^{-4}$. 
Finally, the integrand is at least of order
$Im \lbrace \partial \ln  s_{12}^{(0)} / \partial {\overline X} \rbrace
(d {\overline X} / dt) T_0$, or
 $Im \lbrace \partial \ln s_{12}^{(0)} / \partial {\overline Y} \rbrace
(d {\overline Y} / dt) T_0$, which are both at least of order 
$\eta^{-1} \times \eta  \times \eta^{-4}  \sim  \eta^{-4}$.

\vspace{1. mm}
 We now evaluate the behavior of the noise for large $\eta$. When $\omega \ll \epsilon_F$, 
only the low order Fourier components of $\eta(t)$ are important. 
$\epsilon_1$ and $\epsilon^{\prime}_2$ will be much smaller than $\epsilon_F$.
At $\epsilon_1 = \epsilon^{\prime}_2=0$, 
$s_{11}^{(0)} \simeq - e^{-2i k_F a} \bigl(1 + \, i /{\overline X} \, + o(\eta^{-2}) \bigr)$
 and $s_{22}^{(0)}$ is the same but ${\overline X}$ is replaced by ${\overline Y}$.
 $s_{12}^{(0)} = -(1 / 2 {\overline X} \, {\overline Y}) \,
 \bigl(1 + i \, cotan(2 k_F a)\bigr) + o(\eta^{-3})$.
 As a result, in Eq. (\ref{Sbasic}), 
$s(\epsilon_1,t)^{\dagger}\sigma_z s(\epsilon_2,t) = \sigma_z + o(\eta^{-2})$,
 the same holds for 
$s(\epsilon_2,t^{\prime})^{\dagger} \sigma_z s(\epsilon_1,t^{\prime})$, 
so that the trace is $o(\eta^{-2})$, 
which yields $S$ of order $\eta^{-2}$, at least.

\section{}

In this appendix, we show the result of calculations for the integrals $I_1$ and $I_2$, 
using Eqs. (\ref{thetaparticular}) and (\ref{Tparticular}) for $\theta(x)$ and $T(x)$.
 The integrand of $I_1$ and $I_2$ are non-zero only if
 at least one of $x$ or $x^{\prime}$ is within distance $l$ from an $x_i$, $i=1,2$.
 We shall need the integrals
\begin{eqnarray}
M_0 \, &=& \, \int_0^1 \exp\Bigl({x^2 \over x^2-1}\Bigr) \, dx \, \simeq 0.603,\\
K_1 \, &=& \, \int_{-1}^1 \exp\Bigl({x^2 \over x^2-1}\Bigr) \, (1-x)^{-1} \, dx \simeq 1.207,\\
K_2 \, &=&\, \int_{-1}^1 \! \int_{-1}^1
 \biggl\lbrack \exp\Bigl({x^2 \over x^2 -1}\Bigr) - \exp\Bigl({y^2 \over y^2 -1}\Bigr)\biggr\rbrack^2
\, (x-y)^{-2} dx\, dy \, \simeq \, 2.088, \\
L_1 \,  &=&   \int_{-1}^{1} \! \int_{-1}^1 
\Biggl\lbrace \exp\Bigl({x^2 \over x^2-1}\Bigr) + 
\exp\Bigl({y^2 \over y^2 -1}\Bigr) -
 2 \exp\biggl({1 \over 2} \,\Bigl\lbrack {x^2 \over x^2-1} + {y^2 \over y^2-1}\Bigr\rbrack \biggr)\, 
cos\Bigl({\pi \over 2} (x-y)\Bigr) 
\Biggr\rbrace \, \nonumber \\
\,\,\,&\,& (x-y)^{-2} \, dx\, dy \, \simeq 7.997,\\
L_2 \,  &=&   \int_{-1}^{1} \! \int_{-1}^1 
\Biggl\lbrace \exp\Bigl({2x^2 \over x^2-1}\Bigr) + 
\exp\Bigl({2y^2 \over y^2 -1}\Bigr) -
\exp\Bigl({1 \over 2} \,\lbrack {x^2 \over x^2-1} + {y^2 \over y^2-1}\rbrack \Bigr)\,
\Bigl\lbrack \exp({x^2 \over x^2-1}) + \exp({y^2 \over y^2-1}) \Bigr\rbrack \, \nonumber \\
\,\,\,& \, & \,\,\,\,\, cos\Bigl({\pi \over 2} (x-y)\Bigr) 
\Biggr\rbrace \, (x-y)^{-2} \, dx\, dy \, \simeq \,  6.591.
\end{eqnarray}
 
Then, plugging these values into  (\ref{Q0sint}), (\ref{I1}) and (\ref{I2}),
\begin{eqnarray}
Q_0 \,  &=& \, e (1 - T_1 C_1), \\
S_0 \, &=& \, C_2  \Bigl({\omega \over 2 \pi}\Bigr) \, 
(e-Q_0)  \, \Bigl\lbrack e - C_3(e-Q_0)\Bigr\rbrack,
\end{eqnarray}
with
\begin{eqnarray}
C_1 \, &=& \, M_0 \, \simeq \, 0.603,  \\
C_2 \, &=&\,  {2 \over \pi^2} \, (8 K_1 + 2L_1)/M_0 \, \simeq \, 8.613, \\
C_3 \, &=& \, {L_2 - K_2 \over (L_1 + 4 K_1) M_0} \simeq \, 0.58.
\end{eqnarray}

\section{}

In this appendix, we show the equivalence of Eq. (\ref{noiseMn}) with
 Eqs. (24a), (24b) and (24c) of Ref. \onlinecite{MB04}. There, the noise power 
$P_{\alpha \beta}$ between leads $\alpha$ and $\beta$ was given by
\begin{eqnarray}
P_{\alpha \beta} \,&=&\, 2 {e^2 \over h} \sum_{q=1}^{\infty}
 q \hbar \omega \, C^{sym}_{\alpha \beta,q}(\mu), \\
C_{\alpha \beta, q}^{sym}(E) \,&=&\, {C_{\alpha \beta,q}(E) + C_{\alpha \beta,-q}(E) \over 2}, \\
C_{\alpha \beta,q}(E) \, &=& \, \sum_{\gamma} \sum_{\delta} 
\bigl\lbrack s_{\alpha \gamma}^*(E) s_{\alpha \delta}(E) \bigr\rbrack_q
\bigl\lbrack s_{\beta \delta}^*(E) s_{\beta \gamma}(E) \bigr\rbrack_{-q},
\end{eqnarray}
where $\lbrack A \rbrack_q$ denotes the Fourier transform at frequency
 $q \omega$ of a time dependent quantity $A$. 
In our case, there are only two leads so that indices $\gamma$ and $\delta$ 
are either $1$ or $2$. We are interested
 in $P_{11}$, so we make $\alpha= \beta=1$. Then, inserting the value of the scattering matrix 
elements according to Eq. (\ref{sm}), leads to
\begin{eqnarray}
P_{11} \, =\, 4 e^2 \Bigl({\omega \over 2 \pi}\Bigr) \, 
\Biggl\lbrack 
\sum_{q \geq 1} q 
\biggl(
\vert T_q \vert^2 + {1 \over 2} 
\Bigl\lbrace 
\vert (\sqrt{RT} e^{i \theta})_q\vert^2 + 
\vert (\sqrt{RT} e^{- i \theta})_q\vert^2 
\Bigr\rbrace
\biggr)
\Biggr\rbrack.
\label{resultMB04}
\end{eqnarray}

On the other hand, using Eq.(\ref{noiseMn}), we obtain
\begin{eqnarray}
S \, =\, 8 e^2 
\Bigl({\omega \over 2 \pi}\Bigr) \, 
\Biggl\lbrack 
\sum_{n \geq 1} n 
\biggl(
\vert T_n \vert^2 + {1 \over 2} 
\Bigl\lbrace 
\vert (\sqrt{RT} e^{i \theta})_n\vert^2 + 
\vert (\sqrt{RT} e^{- i \theta})_n\vert^2 
\Bigr\rbrace
\biggr)
\Biggr\rbrack,
\label{resultDGM}
\end{eqnarray}
which is, apart from an overall factor $2$, the same as Eq. (\ref{resultMB04}),
 with $q$ changed to $n$.

\end{document}